\documentclass[aps,prb,twocolumn,amsmath,notitlepage,floatfix,footinbib,superscriptaddress,showpacs, showkeys]{revtex4-1}
\usepackage{graphicx}
\usepackage{wasysym}
\usepackage{amssymb}
\usepackage{mathptmx}
\usepackage[utf8]{inputenc}
\usepackage[english]{babel}
\usepackage[usenames,dvipsnames]{color}
\usepackage{wrapfig}
\usepackage{longtable}
\usepackage{array}
\usepackage{multirow}
\usepackage[colorlinks=true,citecolor=blue,urlcolor=blue,linkcolor=blue,hyperfigures=true]{hyperref}
\usepackage{tikz}
\usepackage{scalefnt}
\usepackage{todonotes}
\usepackage{titletoc}
\usepackage[title,titletoc]{appendix}
\usetikzlibrary{calc}
\usetikzlibrary{arrows}

\newcommand\Let{\mathrel{\mathop:\!\!=}}

\newcommand{\kay}{\mathbf{k}}
\newcommand{\cue}{\mathbf{q}}
\newcommand{\up}{\uparrow}
\newcommand{\down}{\downarrow}
\newcommand{\str}{^{*}}
\newcommand{\da}{^{\dagger}}


\begin{document}
\title{Multiband Dual Fermion Approach to Quantum Criticality in the Hubbard Honeycomb Lattice}  

\author{Daniel Hirschmeier}
\affiliation{I. Institut f\"ur Theoretische Physik, Universit\"at Hamburg, Jungiusstra\ss e 9, D-20355 Hamburg, Germany}
\author{Hartmut Hafermann}
\affiliation{Mathematical and Algorithmic Sciences Lab, Paris Research Center, Huawei Technologies France SASU, 92100 Boulogne-Billancourt, France}
\author{Alexander~I.~Lichtenstein}
\affiliation{I. Institut f\"ur Theoretische Physik, Universit\"at Hamburg, Jungiusstra\ss e 9, D-20355 Hamburg, Germany}

\date{\today}

\begin{abstract}
We study the Hubbard model on the honeycomb lattice in the vicinity of the quantum critical point by means of a multiband formulation of the Dual Fermion approach. 
Beyond the strong local correlations of the dynamical mean field, critical fluctuations on all length scales are included by means of a ladder diagram summation. Analysis of the susceptibility yields an estimate of the critical interaction strength of the quantum phase transition from a paramagnetic semimetal to an antiferromagnetic insulator, in good agreement to other numerical methods. We further estimate the crossover temperature to the renormalized classical regime. Our data imply that, at large interaction strengths, the Hubbard model on the honeycomb lattice behaves like a quantum nonlinear $\sigma$ model, while displaying signs of non-Fermi liquid behavior.
\end{abstract}


\maketitle

\section{Introduction}
The Hubbard model is widely believed to capture some of the most exciting phenomena in strongly correlated electron systems, including unconventional superconductivity~\cite{Dagotto_RMP}, itinerant magnetism, non-Fermi-liquid behavior~\cite{NFL_in_dnf_RMPD} and quantum criticality. The nature of its ground state is determined decisively by the underlying lattice.

Since the discovery of graphene \cite{Graphene_RevModPhys}, the half-filled Hubbard model on the honeycomb lattice has been intensely studied with a variety of techniques, each having their respective advantages and limitations. They nevertheless agree concerning the existence of a quantum critical point at finite value of the interaction strength. 

The application of projective quantum Monte Carlo to the ground state has raised speculations about the existence of a spin-liquid phase at zero temperature \cite{spinliquid}. Later studies revealed that an extrapolation from sufficiently large lattice sizes causes the semimetal-insulator transition and the onset of an antiferromagnetically ordered ground state to coincide\cite{Sorella,PhysRevX.3.031010}. This leaves a vanishingly small window for the existence of a spin liquid. 

Thus far the value of the critical Hubbard interaction $U_c$ for the quantum phase transition from the semimetal to the antiferromagnet has been calculated with a variety of methods, with results ranging from 3.5 to 5 (all values in units of the hopping t). The more recent works almost entirely find values close to 3.8. The introduction section of Ref.~\onlinecite{TPSC} provides a good overview over existing results, while here we only quote the more recent ones, that agree closely and that we consider the most accurate: Large scale projective quantum Monte Carlo (QMC) \cite{Sorella} yields $U_c = 3.869 \pm 0.0013$, while the pinning-field QMC measurements used in Ref.~\onlinecite{PhysRevX.3.031010} result in $U_c = 3.78 \pm 0.001$. The dynamical cluster approximation (DCA) gives $U_c=3.69$ using up to 96 cluster sites \cite{wuandtremblay}. Functional renormalization group investigations \cite{PhysRevLett.100.146404,PhysRevLett.100.156401} and the Variational Cluster approach \cite{VCA} find $U_c = 3.8$, while a study based on the two-particle self-consistent (TPSC) method reports $U_c = 3.79 \pm 0.01$ \cite{TPSC}.

In this paper, we address this problem by means of the Dual Fermion (DF) approach~\cite{PhysRevB.77.033101}. DF belongs to the recently developed important class of methods known as diagrammatic extensions of dynamical mean-field theory~\cite{Rohringer2017}. Their common feature is a perturbative expansion of the self-energy in terms of a dynamical vertex of the underlying DMFT impurity problem. While the self-energy is approximate, correlations are included on all length scales.
Notably, diagrammatic extensions have been shown to recover nonlocal correlation phenomena such as pseudogap formation \cite{Hafermann2009}, formation of extended van Hove singularities \cite{condensation} and (quantum) criticality \cite{PhysRevLett.115.036404} including non mean-field critical exponents \cite{Hirschmeier2015}. The D$\Gamma$A approach, which is similar in spirit to the Dual Fermion approach, has recently been extended to multiorbital systems\cite{MODGA}.

The reasons for applying DF to the Hubbard honeycomb lattice are twofold: firstly, we aim to provide an independent viewpoint on the physics of this model by a method that is quite different in spirit from the ones previously applied to the problem as mentioned above; secondly, we show that DF captures the physics of the quantum critical point and even yields a quantitative estimate of the critical interaction.
To this end, we generalize the ladder DF approach (LDFA)~\cite{Hafermann2009} to account for multiple atoms in the unit cell in a multiband extension. 
Remarkably we find that the resulting method is able to capture the physics of the Hubbard model on honeycomb lattice, which differs qualitatively from that of the square lattice Hubbard model~\cite{Katanin2009,Rost2012,vanLoon2018}. 
This approach allows us in particular to numerically establish the connection between the large interaction limit of the Hubbard model on the honeycomb lattice and the renormalization group treatment of the quantum nonlinear $\sigma$ model as in Ref.~\onlinecite{PhysRevLett.60.1057}.

\section{Model and Method}\label{sec:mod_meth}

\begin{figure}[t]
\begin{center}
\scalebox{1.1}{\tikzset{
        reed/.style={circle,fill=red!100,draw=red!100,minimum size=2mm,inner sep=0mm},
        blck/.style={circle,fill=black!100,draw=black!100,minimum size=2mm,inner sep=0mm},
        impa/.style={circle,fill=black!100,draw=blue!80,minimum size=2mm, inner sep=0mm},
	impb/.style={circle,fill=red!80,draw=blue!80,minimum size=2mm, inner sep=0mm},
    	}
\begin{tikzpicture}
\node (tag1) at (-1.3, 1.0) [] {a)};
\node[line width = 0.5mm] (1A) at (0.0 , 0.0) [impa,label=above:A] {}; 
\node[line width = 0.5mm] (2B) at (0.5,0.866025) [impb,label=above:\textcolor{red}{B}] {}; 
\node[line width = 0.5mm] (3B) at (0.5,-0.866025) [impb,label=above:\textcolor{red}{B}] {};
\node[line width = 0.5mm] (1B) at (-1., 0.) [impb,label=above:\textcolor{red}{B}] {}; 
\node[line width = 0.5mm] (2A) at (1.5,0.866025) [impa,label=above:A] {};
\node[line width = 0.5mm] (3A) at (1.5,-0.866025) [impa,label=above:A] {};
\draw[-,blue!80, line width=0.5mm] (1A) -- (1B);
\draw[-,blue!80, line width=0.5mm] (2B) -- (2A);
\draw[-,blue!80, line width=0.5mm] (3B) -- (3A);
\draw[-] (1A) -- (2B);
\draw[-] (1A) -- (3B);
\node (tag2) at (2.7, 1.0) [] {b)};
\node[line width = 0.5mm] (1a) at (4.0 , 0.0) [impa,label=above:A] {}; 
\node[line width = 0.5mm] (2b) at (4.5,0.866025) [impb,label=above:\textcolor{red}{B}] {}; 
\node[line width = 0.5mm] (3b) at (4.5,-0.866025) [impb,label=above:\textcolor{red}{B}] {};
\node[line width = 0.5mm] (1b) at (3., 0.) [impb,label=above:\textcolor{red}{B}] {}; 
\draw[-] (1a) -- (1b);
\draw[-] (1a) -- (2b);
\draw[-] (1a) -- (3b);
\end{tikzpicture}}
\caption{\label{fig:cluster_symmetry} 
a) In a cluster approach lattice symmetry is broken since self-energies obtained from the cluster (thick lines) or perturbatively (thin lines) treat a priori equivalent bonds unequally.
b) In the multiband approach all nonlocal self-energies are calculated on equal footing.}
\end{center}
\end{figure}
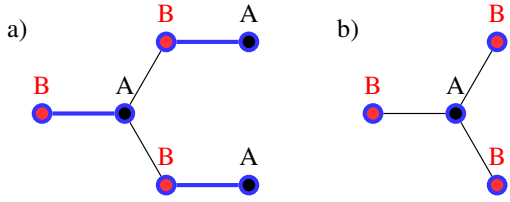

We formulate the problem for two atoms in the unit cell adapted to the honeycomb lattice, although the extension to an arbitrary number of atoms is straightforward. For each atom one in general has to solve an impurity problem, similarly to real-space DMFT (RDMFT) \cite{RDMFT}. For the honeycomb lattice one has two equivalent impurity problems owing to the equivalence of two sublattices.
In DMFT the inter-site self-energy is zero, whereas in DF it is included diagrammatically. In principle it is possible to start from a cluster formulation and to include diagrams beyond DCA~\cite{Yang2011a} or CDMFT~\cite{Hafermann2008}, which however breaks lattice symmetries artificially depending on the choice of the cluster problem (see Fig.~\ref{fig:cluster_symmetry}).

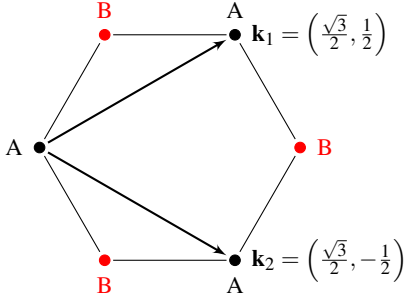
\begin{figure}[t]
  \scalebox{1.0}{\begin{tikzpicture}
\node (oneA) at (0.0 , 0.0) [label=left:A] {}; 
\filldraw[black] (oneA) circle (2pt);
\node (twoA) at (3*0.866025, 3*0.5) [label=above:A, label=right:{$\mathbf{k}_{1}=\left(\frac{\sqrt{3}}{2},\frac{1}{2}\right)$}] {}; 
\filldraw[black] (twoA) circle (2pt); 
\node (threeA) at (3*0.866025, 3*-0.5) [label=below:A, label=right:{$\mathbf{k}_{2}=\left(\frac{\sqrt{3}}{2},-\frac{1}{2}\right)$}] {};
\filldraw[black] (threeA) circle (2pt);
\node (oneB) at (3*1.154700538, 0.) [label=right:\textcolor{red}{B}] {}; 
\filldraw[red] (oneB) circle (2pt);
\node (twoB) at (3*0.28867513, 3*0.5) [label=above:\textcolor{red}{B}] {}; 
\filldraw[red] (twoB) circle (2pt);
\node (threeB) at (3*0.28867513, 3*-0.5) [label=below:\textcolor{red}{B}] {}; 
\filldraw[red] (threeB) circle (2pt);

\draw[->,>=latex',thick] (oneA) -- node [below] {} (twoA);
\draw[->,>=latex',thick] (oneA) -- node [below] {} (threeA);
\draw[-] (oneA) -- (twoB);
\draw[-] (twoB) -- (twoA);
\draw[-] (twoA) -- (oneB);
\draw[-] (oneB) -- (threeA);
\draw[-] (threeA) -- (threeB);
\draw[-] (threeB) -- (oneA);

\end{tikzpicture}}
  \caption{Sketch of the bipartite honeycomb lattice consisting of sublattices A and B.}
  \label{fig:honey}
\end{figure}

For the Hubbard model on a lattice with multiple atoms per unit cell the action is given by
\begin{align}
S[c\str,c] = &-\sum_{\nu,\kay,A,B,\sigma} c\str_{\nu\kay A\sigma}\left[ (i\nu+\mu)\delta_{AB} -\hat{\epsilon}(\kay)_{AB} \right]c_{\nu\kay B\sigma} \notag \\ 
&+ U \sum_{\omega,i,A}n_{\omega i A \up} n_{-\omega i A \down}, \label{eq:hub_action}
\end{align}
where $\nu$ ($\omega$) represents fermionic (bosonic) Matsubara frequencies and $\kay$ the lattice momentum, $\beta=1/T$ is the inverse temperature and $\mu$ the chemical potential. We assume that the Coulomb repulsion is restricted to on-site terms of strength $U$.
The lattice is divided into sublattices as shown in Fig.~\ref{fig:honey}. The sum over sublattice indices $A,B$ runs over all $N$ atoms in the unit cell. Small Latin indices label the sites on the Bravais lattice (conjugate variables to the lattice momenta) and $\sigma=\up,\down$ represents the spin projection. We generally drop spin indices on single-particle quantities as we only consider the SU(2) symmetric case.
The dispersion $\hat{\epsilon}(\kay)$ is an $N\times N$ matrix in sublattice indices. Its off-diagonal elements account for the hopping between sublattices. Here and in the following we indicate matrix-valued functions with a caret.

\subsection{Dual Fermions}

The DF approach is a means to include nonlocal correlations beyond the dynamical mean-field level diagrammatically. 
For a derivation and introduction to the method, we refer the reader to a recent review of diagrammatic extensions of DMFT~\onlinecite{Rohringer2017}.
In this paper, we generalize the approach to the multiband case, more specifically to multiple atoms within a unit cell. We briefly outline the derivation and discuss the structure of the DF approach specific to the multiband case. Further technical details of the derivation and the approach can be found in the Appendix.

In the DF approach, the lattice problem is reformulated in terms of auxiliary, so-called dual fermions, such that the case of noninteracting dual fermions corresponds to DMFT.
To treat strong local correlations, we first introduce Anderson impurity problems for each site in both sublattices and express the lattice action in terms of these as follows:
\begin{equation}
S[c\str,c] = \sum_{i A} S_{\text{imp}}[c\str_{i A},c_{i A}]-\!\!\!\sum_{\nu,\kay,A,B,\sigma}\!\! c\str_{\nu\kay A\sigma}\left[ \hat{\Delta}_{\nu}-\hat{\epsilon}_{\kay} \right]_{A B}c_{\nu\kay B}, \label{A_hub_action_IV}
\end{equation}
where the hybridization matrix of the impurity problem $\hat{\Delta}$ is diagonal in sublattice space, $\hat{\Delta}_{AB}=\Delta_{A}\delta_{AB}$. 

The impurities are coupled through the second term in the equation above. They are decoupled through a Hubbard-Stratonovich transformation [the explicit form is given in~\eqref{eq:A_HST_II}], which introduces new Grassmann fields $f$,$f^*$.
These dual fermions can be thought of as mediating the coupling between the impurities on different sites of the Bravais lattice. 

As shown in the Appendix, the physical fermion variables can formally be integrated out exactly. The result is that the action is exactly mapped onto an equivalent one in terms of dual fermions,
\begin{align}
\tilde{S}[f^{*},f] &=-\sum_{\nu\kay AB\sigma} f^{*}_{\nu\kay A\sigma}\left[\tilde{G}^{(0)}_{\nu\kay}\right]_{AB}^{-1}f_{\nu\kay B\sigma} + \sum_i \tilde{V}[f^{*}_i,f_i], \notag\\
\hat{\tilde{G}}^{(0)}_{\nu\kay} &=\left[\hat{g}^{-1}_\nu+(\hat{\Delta}_\nu-\hat{\epsilon}_\kay)\right]^{-1}-\hat{g}_\nu,\notag\\
V[f\str,f] &=-\frac{1}{4}\sum_A\sum_{\alpha\beta\gamma\delta}\gamma^{A}_{\alpha\beta\gamma\delta}f\str_{A\alpha}f_{A\beta}f\str_{A\gamma}f_{A\delta} + \ldots ,
\end{align}
where dual quantities are marked by a tilde and Greek letters denote compound indices, $\alpha=\{\nu\sigma\}$.
Note that since the hybridization and Coulomb interaction are local, the impurity vertex is diagonal in sublattice indices as well,  $\gamma^{ABCD}=\gamma^{AAAA}\delta_{AB}\delta_{BC}\delta_{CD}$, and we abbreviate $\gamma^{A}\equiv\gamma^{AAAA}$. 

Details of this transformation can be found in the Appendix. 
The underlying idea is that a perturbative solution of the problem in terms of the bare dual Green's function $\hat{\tilde{G}}^{(0)}_{\nu\kay}$ and dual interaction $V[f\str,f]$ contains the correlated non-perturbative local physics from the start (at zeroth order). 
The nonlocal correlations are presumably weaker than the local ones and are therefore treated diagrammatically.
The elements of the new action can be obtained numerically from the solution of the impurity model, for which established and accurate techniques exist~\cite{Gull2011}.

While for graphene the sublattices are equivalent, symmetry breaking between sublattices can be induced locally in our approach for example through a sublattice-dependent chemical potential or Coulomb interaction, or via the dispersion matrix (consider a bilayer honeycomb lattice with layer-dependent hopping amplitudes).

\subsection{Ladder approximation}

\begin{figure}[t!]
\begin{center}
{\scalefont{1.5}
\scalebox{0.55}{\input{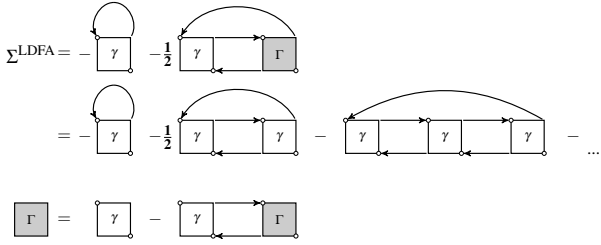}}}
\caption{\label{fig:sde} Diagrammatic representation of the Schwinger-Dyson equation (first line), the three lowest order contributions to the dual self-energy within the ladder Dual Fermion approach (second line) and the Bethe-Salpeter equation. The second line follows from the first line if the vertex is expanded according to the Bethe-Salpeter equation.}
\end{center}
\end{figure}

To describe the physics in the vicinity of the quantum critical point, we need an approximation that includes long-range fluctuations. Physically we expect (magnetic) particle-hole fluctuations to dominate. The corresponding diagrams are depicted in Fig.\ref{fig:sde}. 
The approximation is akin to the fluctuating exchange approximation (FLEX)\cite{FLEX}, albeit in dual space. The Bethe-Salpeter equation (BSE) is given by (all quantities such as Green's functions in this section are dual by default)
\begin{equation}\label{eq:bse}
\hat{\tilde{\Gamma}}_{\nu\nu'\omega\cue}^{c} =\hat{\tilde{\boldsymbol{\gamma}}}_{\nu\nu'\omega}^{c}-\frac{1}{\beta}\sum_{\nu''}\hat{\tilde{\boldsymbol{\gamma}}}_{\nu\nu''\omega}^{c}\hat{\tilde{\boldsymbol{\chi}}}^{(0)}_{\nu''\omega\cue}\hat{\tilde{\Gamma}}_{\nu''\nu'\omega\cue}^{c},
\end{equation}
where $\hat{\tilde{\Gamma}}^{c}$ is the dual renormalized lattice vertex and $\hat{\tilde{\gamma}}^{c}$ the reducible impurity vertex in spin and charge channels $c=\text{sp},\text{ch}$. The latter vertex is taken as an approximation for the dual irreducible vertex. The charge and spin components of the vertices are given by $\Gamma^{\text{sp(ch)}}=\Gamma^{\up\up\up\up}\stackrel{(+)}{-}\Gamma^{\up\up\down\down}$.
$\hat{\tilde{\chi}}^{(0)}$ is the dual particle-hole bubble,
\begin{align}
\tilde{\chi}^{(0)AB}_{\nu\omega\cue} & = \frac{1}{N}\sum_\kay \tilde{G}^{AB}_{\nu+\omega\kay+\cue}\tilde{G}^{BA}_{\nu\kay},
\label{eq:ph_bubble2}
\end{align}
where we have used the shorthand notation $\tilde{\chi}^{AB}\equiv\tilde{\chi}^{AABB}$. Note that only terms of this form contribute because the local vertex is diagonal in the sublattice indices.

\twocolumngrid

The matrix structure of the quantities in the BSE in sublattice space is as follows:
\begin{align}
\hat{\tilde{\Gamma}}_{\nu\nu'\omega\cue}^{c}&=
\begin{pmatrix}
\tilde{\Gamma}^{AA c}_{\nu\nu'\omega\cue} & \tilde{\Gamma}^{AB c}_{\nu\nu'\omega\cue} \\
\tilde{\Gamma}^{BA c}_{\nu\nu'\omega\cue} & \tilde{\Gamma}^{BB c}_{\nu\nu'\omega\cue}  \\
\end{pmatrix},\notag\\
\hat{\tilde{\boldsymbol{\chi}}}^{(0)}_{\nu\omega\cue} & =\begin{pmatrix}
\tilde{\chi}^{(0)AA}_{\nu\omega\cue} & \tilde{\chi}^{(0)AB}_{\nu\omega\cue} \\
\tilde{\chi}^{(0)BA}_{\nu\omega\cue} & \tilde{\chi}^{(0)BB}_{\nu\omega\cue} \\
\end{pmatrix},\notag\\
\hat{\tilde{\boldsymbol{\gamma}}}_{\nu\nu'\omega}^c &=\begin{pmatrix}
\tilde{\gamma}^{A c}_{\nu\nu'\omega} & 0 \\
0 & \tilde{\gamma}^{B c}_{\nu\nu'\omega}\\
\end{pmatrix}\label{eq:matrices}.
\end{align}
Note that the off-diagonal components of the lattice vertex stem solely from the bubble, which mixes the sublattice components.
The solution of the BSE is obtained by inverting $\hat{\tilde{\Gamma}}^{-1}=[\hat{\tilde{\gamma}}^{-1}+\hat{\tilde{\chi}}^{(0)}]^{-1}$, for each channel and each value of the bosonic frequency $\omega$ up to the cutoff. Correspondingly the matrices should be viewed as matrices in a combined frequency and sublattice index.

From the renormalized two-particle vertex we obtain the self-energy in the ladder approximation by means of the dual version of the Schwinger-Dyson equation (SDE). It is depicted in Fig.~\ref{fig:sde} and reads
\begin{figure*}[t]
\begin{center}
\includegraphics[width=1.5\columnwidth]{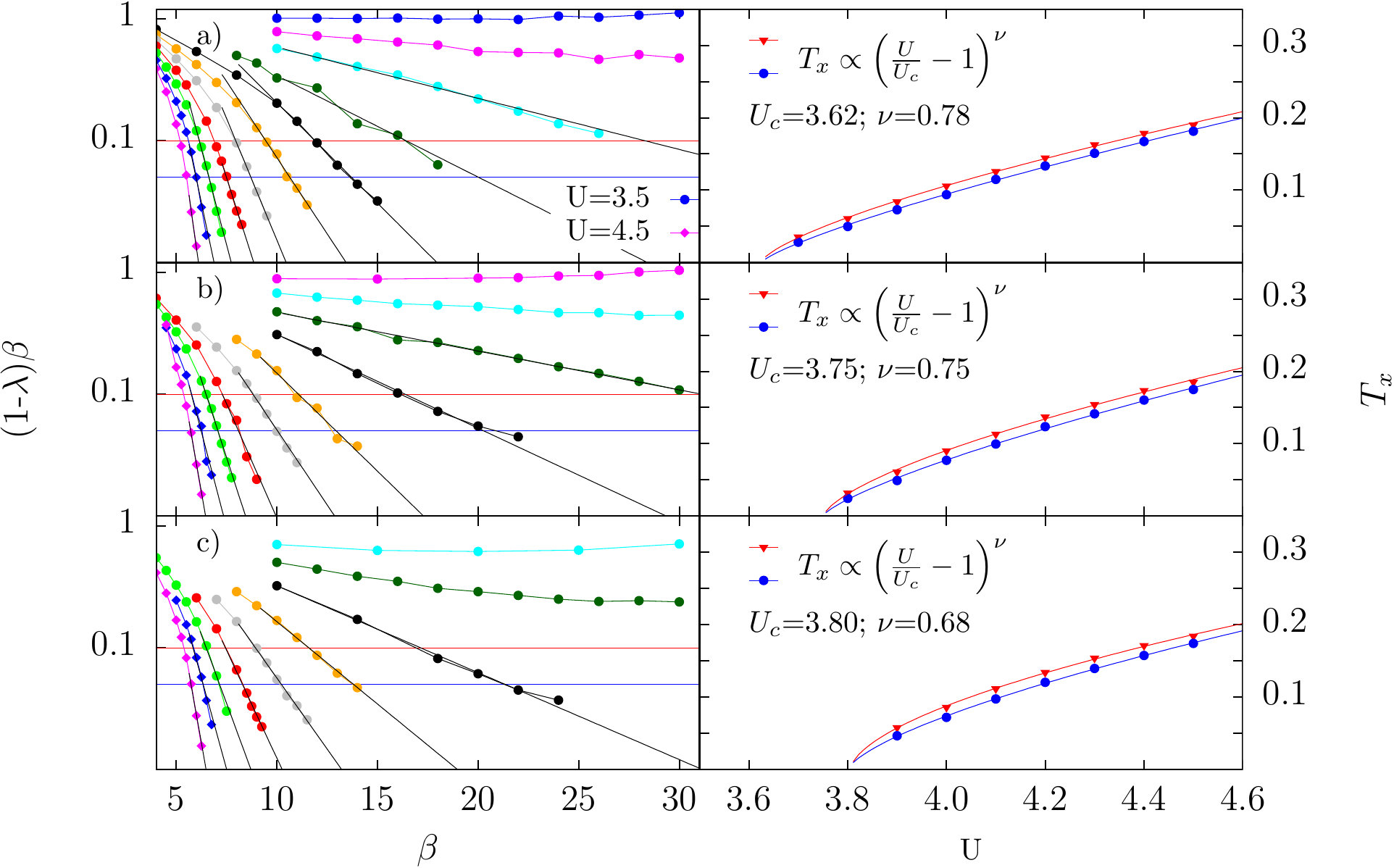}\end{center}\vspace*{-1.5em}
\caption{\textit{Left column}: Log scale plot of $(1-\lambda)\beta$ against inverse temperature $\beta$ for different numbers of bosonic frequencies entering the BSE a) $N_W=1$ (first row), b) $N_W=4$ (second row), c) $N_W=16$ (third row) and various values of $U$ in steps of $0.1t$. For $U=U_c$, $(1-\lambda)\beta$ is nearly constant and decays exponentially for $U>U_c$. Black solid lines are exponential fits to the data. \textit{Right column}: Estimate for the crossover temperature $T_x$ to the renormalized classical regime against interaction $U$. Data points are obtained from intersections of fits to $(1-\lambda)\beta$ with constants 0.1 (red) and 0.05 (blue) (horizontal lines in left column). Solid lines are fits $T_x\propto(U/U_c-1)^{\nu}$, with $U_c$ and $\nu$ as fit parameters. Fit results for $U_c$ agree well with $U_c$ from left column, while results for $\nu$ are in good agreement with the literature as they are close to $\nu=0.7$. The values for $U_c$ and $\nu$ shown in the plot correspond to the blue line fit.}
\label{fig:susc_eig_log}
\end{figure*}
\begin{align}
\hat{\tilde{\Sigma}}^{AB}_{\nu\kay} = -\frac{T}{N}\sum_{\nu'} &\gamma^{A \text{ch}}_{\nu\nu'\omega=0}G^{AA}_{\nu'}\delta_{AB} \notag\\
 -\frac{1}{2}\frac{T^{2}}{N^{2}}\sum_{\nu'\omega\cue C} &\gamma^{A \text{ch}}_{\nu'\nu\omega}\chi^{(0)CA}_{\nu'\omega\cue}G^{AB}_{\nu+\omega\kay+\cue}\left[2\Gamma^{BC \text{ch}}_{\nu\nu'\omega\cue}-\gamma^{B \text{ch}}_{\nu\nu'\omega}\delta_{BC} \right] \notag \\
 -\frac{3}{2}\frac{T^{2}}{N^{2}}\sum_{\nu'\omega\cue C} &\gamma^{A \text{sp}}_{\nu'\nu\omega}\chi^{(0)CA}_{\nu'\omega\cue}G^{AB}_{\nu+\omega\kay+\cue}\left[2\Gamma^{BC \text{sp}}_{\nu\nu'\omega\cue}-\gamma^{B \text{sp}}_{\nu\nu'\omega} \delta_{BC}\right].
 \label{eq:sde}
\end{align}
The factor $3/2$ stems from the three-fold degeneracy of the spin-1 bosonic excitations in the spin channel.

\subsection{Calculation procedure}

The numerical calculation proceeds as follows. 
We first perform DMFT iterations to obtain the starting point of our perturbation theory.
Since DMFT corresponds to non-interacting dual fermions, this can be achieved by imposing a self-consistency condition on the bare local dual Green's function~\cite{PhysRevB.77.033101}, 
\begin{equation}
\sum_{\kay} \hat{\tilde{G}}^{AA}_{\nu\kay}=0.\label{eq:sc}
\end{equation}
The bare dual Green's function can be written $\hat{\tilde{G}}^{(0)}_{\nu\kay}=\nolinebreak \hat{G}^{DMFT}_{\nu\kay}-\hat{g}_{\nu}$. Eq.~\eqref{eq:sc} is thus equivalent to the self-consistency condition in DMFT as long as no diagrams are taken into account. Our results for the honeycomb lattice can therefore be interpreted as diagrammatic corrections to the DMFT results of Ref.~\onlinecite{Tran2009}.
Once converged, we compute the vertex of the impurity model(s) and evaluate the dual self-energy according to Eqs.~\eqref{eq:bse} and~\eqref{eq:sde}.
The Green's function is computed from the dual Dyson equation
\begin{equation}
\hat{\tilde{G}}=\left[\hat{\tilde{G}}^{-1}_0-\hat{\tilde{\Sigma}}\right]^{-1}.
\end{equation}
The BSE, SDE and Dyson's equation form a nonlinear set of equations which we solve self-consistently.
We then update the impurity hybridization with the goal to fulfill the self-consistency condition~\eqref{eq:sc} for the full dual Green's function.

Finally, after convergence, the lattice Green's function is readily obtained via
\begin{equation}\label{eq:dual_lattice_relationI}
\hat{G}=\left[\left(\hat{g}+\hat{g}\hat{\tilde{\Sigma}}\hat{g} \right)^{-1}+\hat{\Delta}-\hat{\epsilon} \right]^{-1}.
\end{equation}
To evaluate the generalized susceptibility tensor we first transform the renormalized dual lattice vertex to the corresponding physical lattice vertex using Eq.~\eqref{eq:A_dual_vertex_relation}. The generalized susceptibility tensor and the lattice vertex are related via
\begin{align}\label{eq:gen_susc_ten}
&\chi^{AB\,\sigma\sigma'}_{\omega\cue} \Let\langle n_{\omega\cue A\sigma} n_{-\omega-\cue B\sigma'} \rangle=-\frac{T}{N}\sum_{\nu\kay} G^{AB}_{\nu+\omega,\kay+\cue}G^{BA}_{\nu\kay}  \notag\\
&+ \frac{T^2}{N^{2}}\sum_{\nu\nu'\kay\kay'A'B'}G^{AA'}_{\nu+\omega,\kay+\cue}G^{A'A}_{\nu\kay} \ \Gamma^{A'B'\,\sigma\sigma'}_{\nu\nu'\omega,\cue} \ G^{BB'}_{\nu'\kay'}G^{B'B}_{\nu'+\omega,\kay'+\cue}.
\end{align}
The generalized susceptibility facilitates the analysis of phase transitions in multiband systems, since in general the order parameter is unknown and may have a complicated structure. The order parameter for multiband systems can be deduced from the leading eigenvalue of the generalized susceptibility tensor as demonstrated in Ref.~\onlinecite{LewinVolker}.

\section{Results}\label{sec:res}
We now turn to the Hubbard model on the honeycomb lattice. The Hamiltonian is given by
\begin{align}
H&=\sum_{\kay AB} c\da_{\kay,A} \hat{\epsilon}(\kay)_{AB} c_{\kay,B}+U\sum_{i}(n_{i A\up}n_{i A\down}+n_{i B\up}n_{i B\down}),\notag\\
\hat{\epsilon}_\kay &=\begin{pmatrix} 0 & -tf(\mathbf{k}) \\ -tf^{*}(\mathbf{k}) & 0\end{pmatrix}, \notag\\
f(\kay) &=e^{-i( \frac{a}{\sqrt{3}} k_y)}+2e^{i (\frac{a}{2\sqrt{3}} k_y)} \cos\left( \frac{a}{2} k_x \right).
\end{align}
We use the nearest-neighbor hopping amplitude $t$ as the energy unit and set the lattice constant $a=1$. The impurity problem is solved using a hybridization expansion quantum Monte Carlo algorithm~\cite{Hafermann2013} with improved estimators for the vertex function~\cite{spinfreeze}. We employ a finite cutoff of the Matsubara frequencies of the impurity Green's function $g_\nu$ and two-particle vertex $\gamma_{\nu\nu'\omega}$ with $|\nu|\leq (2n_1+1)\pi T$ as well as $|\nu|,|\nu'| \leq(2n_2+1)\pi T$ and $|\omega|\leq 2 m\pi T$. We choose $n_1$=128 and $n_2$=64. The single-particle Green's function is already well in its tail behavior described by the first few powers of $1/(i\nu_n)$ for this value of $n_1$. We further checked that changes in the dual self-energy are negligible when the cutoff $n_2$ is increased. In order to determine the quantum critical point (QCP) the cutoff $m$ for the bosonic frequency and the lattice size are varied (see below).
We checked numerically that in the weak coupling limit $U\leq 1$ our implementation gives results equal to FLEX.
\begin{figure*}[t]
\begin{center}
\includegraphics[width=1.5\columnwidth]{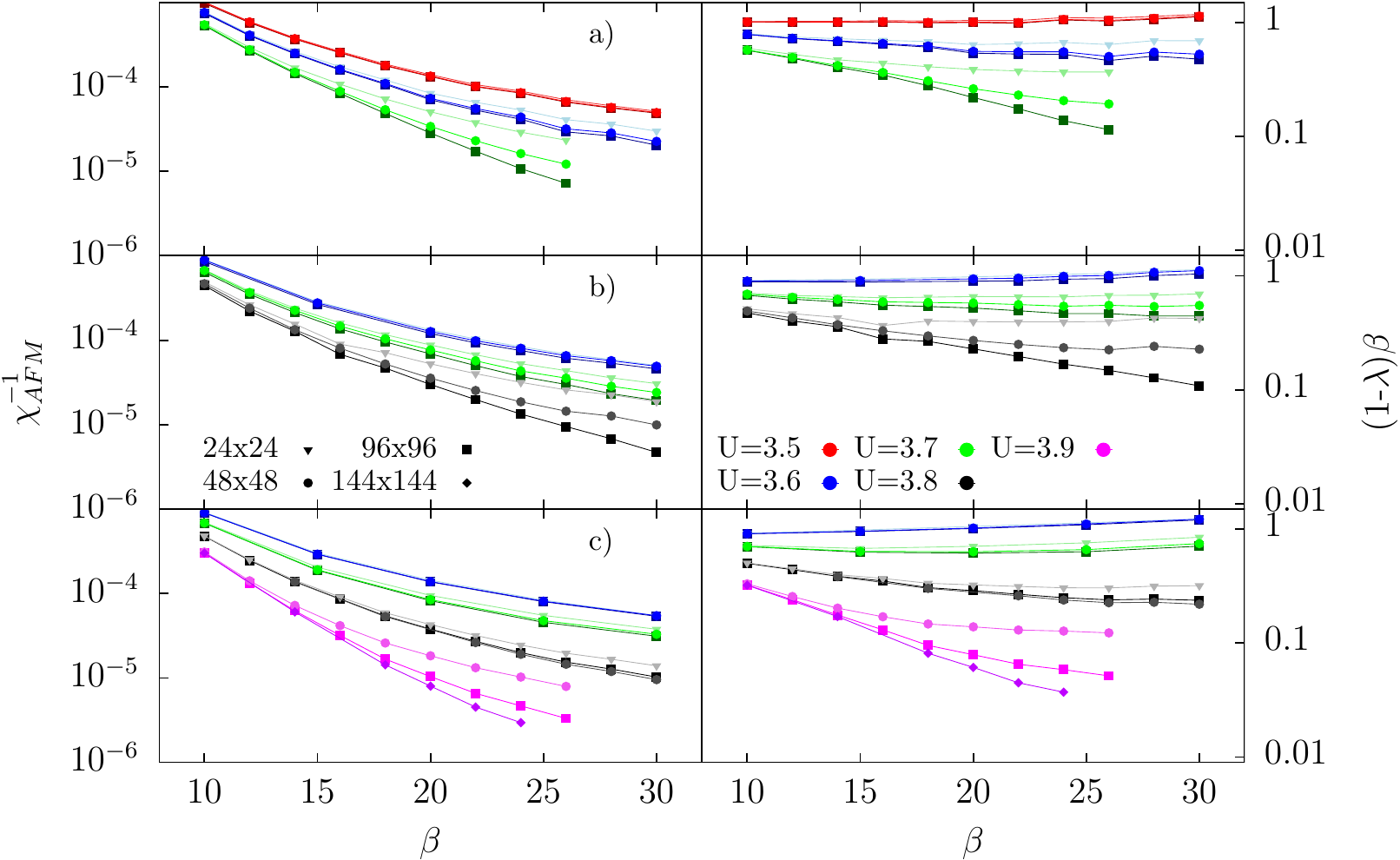}\end{center}\vspace*{-1.5em}
\caption{Size dependence of the inverse AFM susceptibility (left column) and $\tilde{\lambda}=(1-\lambda)\beta$ (right column) for 
a) $N_W=1$ b) $N_W=4$ and c) $N_W=16$ bosonic frequencies. Different point types indicate sizes $24\times 24$ (triangles), $48\times 48$ (squares), $96\times 96$ (bullets), $144\times 144$ (diamonds). Different values of U are indicated by different colors according to the legend in the mid panel. Points are connected with lines as a guide to the eye. For $U>U_c$ these quantities are linear for sufficient lattice sizes in the log scale plot and have a clear size dependence. For $U<U_c$ curves for different sizes are almost on top of one another and $\tilde{\lambda}$ approaches a constant. According to this plot we estimate the critical couplings to be a) $U_c=3.6\pm 0.1$ b) $U_c=3.7\pm 0.1$ c) $U_c=3.8\pm 0.1$, reproducing the results obtained from the fit to $T_x$ of Fig.~\ref{fig:susc_eig_log} rather well.}
\label{fig:sizes}
\end{figure*}
\subsection{Quantum Critical Point}\label{sec:qcp}

We investigate the model in the vicinity of the critical point. As $U$ is increased beyond $U_c$, the ground state changes from a semimetal to an antiferromagnet.
In order to determine $U_c$ we analyze the leading eigenvalue of the matrix $\hat{M}=T\hat{\gamma}\hat{\chi}$, where $\hat{\gamma}$ and $\hat{\chi}$ are defined in \eqref{eq:matrices}. Matrices with leading eigenvalues of $\lambda=1$ are outside of the convergence radius of the geometric series and indicate the divergence of the ladder series due to a phase transition. Hence $1-\lambda$ is a useful measure for the criticality of the system.

At half-filling and large $U$ the Hubbard model can be mapped to an antiferromagnetic Heisenberg model, which can be described by a quantum nonlinear $\sigma$ model in two dimensions in the long-wavelength, low-temperature limit \cite{PhysRevB.39.2344}. In the renormalized classical regime the correlation length scales as $\xi^{-1}\propto 1/\beta\exp(-2\pi\rho_s\beta)$, where $\rho_s$ is the ground state spin stiffness. At the quantum critical point $\rho_s=0$ and therefore $\xi^{-1}\propto 1/\beta$ \cite{PhysRevLett.60.1057}.
The log scale plots on the left hand side of Fig.~\ref{fig:susc_eig_log} reveal that $1-\lambda$ exhibits the same behavior as $\xi^{-1}$ since $(1-\lambda)\beta$ is a constant close to $U=U_c$, while for $U>U_c$ $(1-\lambda)\beta$ decays exponentially. Note that the rate of decay increases as $U$ increases. In order to estimate the crossover temperature $T_x$ where the system enters the renormalized classical regime, the correlation length $\xi$ has been equated to the thermal wavelength of spin waves in Ref.~\onlinecite{PhysRevLett.60.1057} or free electrons in Ref.~\onlinecite{TPSC}. Here, using that $1-\lambda$ is proportional to the inverse correlation length, we can estimate $T_x$ by requiring $(1-\lambda)/T_{x}=C$, where the constant $C$ is chosen such that the exponential decay of $1-\lambda$ is evident from the data for $U>U_c$ and the system has already entered the renormalized classical regime. In particular, we choose $C=0.1$ and 0.05 (see red and blue lines on left hand side of Fig.\ref{fig:susc_eig_log}). Power-law fits to the data $T_x\propto(U/U_c-1)^\nu$ yield good agreement with other numerical methods for the critical coupling $U_c$. We find $\nu=0.7-0.8$ which is close to the value $\nu=0.7$ for the Heisenberg model in 2+1 dimensions~\cite{PhysRevLett.60.1057,Troyer1997}.
\begin{figure*}[t]
\begin{center}
\includegraphics[width=1.5\columnwidth]{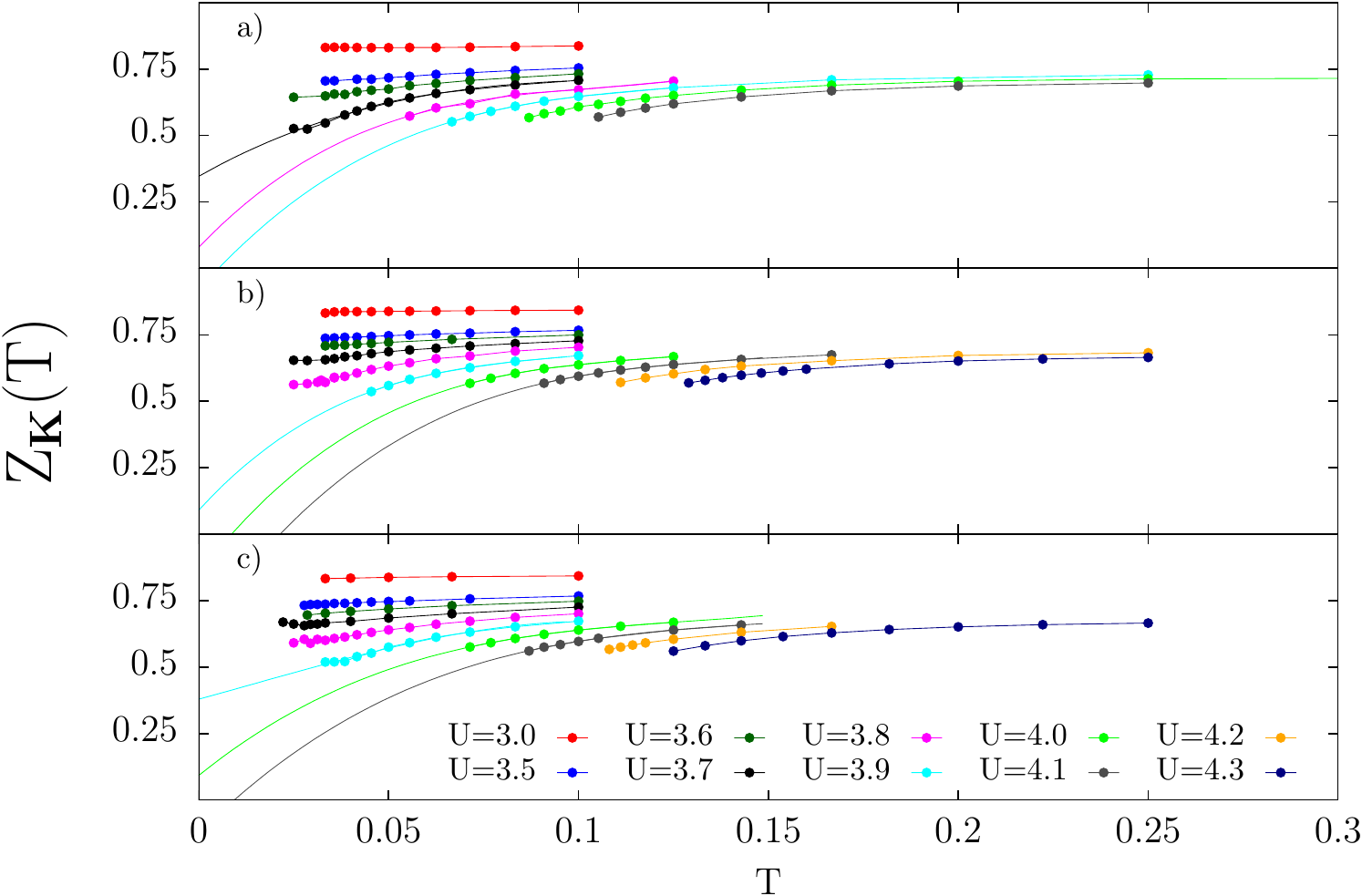}\end{center}\vspace*{-1.5em}
\caption{Quantity $Z_\kay(T)$ defined in \eqref{eq:finite_qpw} at the Dirac point $\kay=\mathbf{K}$ for different values of Hubbard interaction $U$ against temperature according to LDFA for a) $N_W=1$ b) $N_W=4$ and c) $N_W=16$ bosonic frequencies. Polynomial extrapolation is used to determine the value of the Hubbard interaction $U_{SMIT}$ for which the quasiparticle weight vanishes at $T=0$ and the system becomes insulating. The resulting values a) $U_{SMIT}=3.9$ b) $U_{SMIT}=4.0$ c) $U_{SMIT}=4.1$ trail the critical $U_c$ by 0.3t.}
\label{fig:qpw_finite}
\end{figure*}

Analysis of the generalized susceptibility tensor confirms that the system has a tendency to AFM order. In Fig.~\ref{fig:sizes} we plot the inverse of the AFM susceptibility, which is given by  $\langle (S^{z}_{A}-S^{z}_{B})^{2}\rangle_{\omega=0\cue=0}$, with $S^{z}_{A}=n_{A\up}-n_{A\down}$ and $A\neq B$, for different lattice sizes and values of $U$. The data shows no tendency to AFM ordering at finite temperature, as required by the Mermin-Wagner theorem. The susceptibility and leading eigenvalue are size dependent for interaction values corresponding to an AFM ground state\cite{PhysRevLett.115.036404} which is not the case for a semimetallic ground state. As $U$ is decreased starting from $U>U_c$, we can see that the susceptibility and leading eigenvalue become less size dependent and less divergent as the ground state changes from AFM to a semimetal. We define the lower bound for the critical coupling $U_{c1}$ as the value of $U$ for which the according curves are virtually on top of one another, and the upper bound $U_{c2}$ as the value for which the size dependence is evident and $(1-\lambda)\beta$ is a straight line in the log scale plot. We take the estimate for $U_c$ to be their average, ranging from 3.6 to 3.8 in units of the hopping for $N_W=1$ to $N_W=16$ respectively, which gives good agreement with $U_c$ drawn from the fits in Fig.~\ref{fig:susc_eig_log}. For practical numerical purposes it is important to note that using a single bosonic frequency is sufficient to get a good estimate of $U_c$ and $T_x$, while for $N_W=16$ we had to go to lattice sizes of $144\times144$ to recover a straight line in the log scale plot for $U>U_c$. Note that as one increases the number of bosonic frequencies accounted for in the LDFA simulations, the value for the critical coupling $U_c$ is shifted to larger values, as more long-ranged spin fluctuations destroy the AFM order. We found that this saturates and the approach converges for $N_W=16$ bosonic frequencies for the model under consideration.\\
\subsection{Semimetal Insulator Transition}\label{sec:smit}
Being restricted to finite temperatures, we obtain information about the conducting properties of the ground state using polynomial extrapolations of the quantity
\begin{equation}\label{eq:finite_qpw}
Z_\kay(T)=\left[1-\frac{{\rm Im}\hat{\Sigma}_\kay(\pi T)}{\pi T}\right]^{-1},
\end{equation}
evaluated at the Dirac point $\kay=\mathbf{K}$, to the limit $T\rightarrow 0$, which equals the quasiparticle renormalization factor\cite{wuandtremblay}. As the self-energy in our problem is matrix-valued, we consider the trace of Eq.~\eqref{eq:finite_qpw}. 

The corresponding data, again for different number of bosonic frequencies, is shown in Fig.~\ref{fig:qpw_finite}. While projective QMC calculations with size extrapolation \cite{Sorella,PhysRevX.3.031010} indicate that the interaction strengths marking the onset of the AFM phase ($U_c$) and the opening of the single-particle gap ($U_{SMIT}$) coincide, extrapolation of our data instead consistently gives $U_{SMIT}-U_{c}\simeq0.3t>0$, regardless of the number of bosonic frequencies. Therefore our data contains no hint towards the existence of a spin liquid phase. Due to the extrapolation over a rather wide temperature range it is difficult to draw final conclusions whether $U_{SMIT}$ is actually different from $U_c$. We note however that our results from Fig.~\ref{fig:qpw_finite} c) are consistent with those reported in Ref.~\onlinecite{wuandtremblay}.

\begin{figure*}[t]
\begin{center}
\includegraphics[width=1.7\columnwidth]{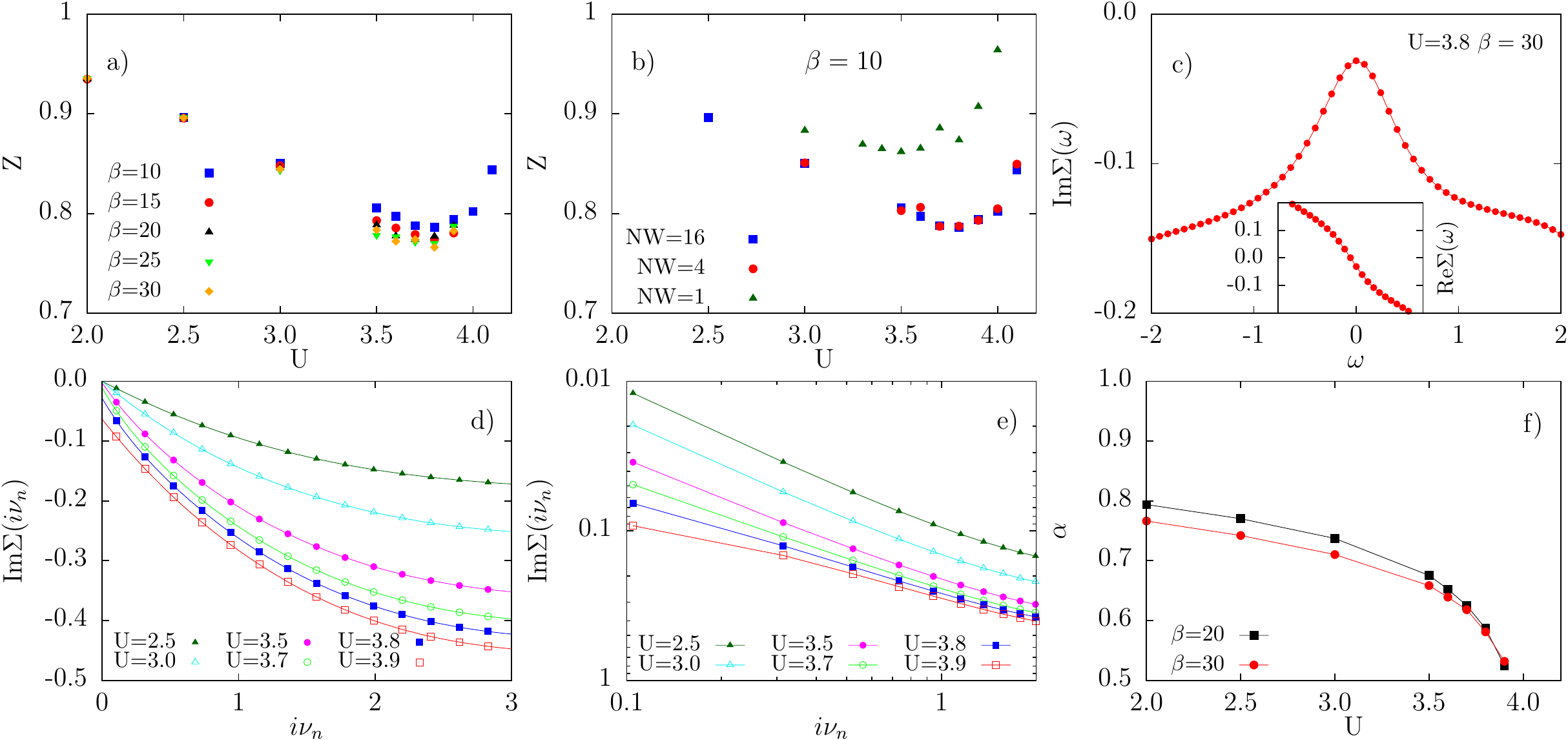}\end{center}\vspace*{-1.5em}
\caption{Analysis of the finite temperature quasiparticle weight (top panels) and self-energy at the Dirac-point (bottom panels). \textit{Top left panel} a) Quasiparticle weight Z from Eq.~\eqref{eq:zeee} versus the Hubbard interaction for different temperatures ($\beta= 10 - 30$) using $N_W=16$. $Z$ exhibits an unphysical increase with U for $U>3.8$. \textit{Top mid panel} b) This behavior is seen as well for $N_W=1$ (green triangles $U>3.5$) and $N_W=4$ (red circles) ($\beta=10$). \textit{Top right panel} c) Example of the imaginary part of the self-energy ${\rm Im}\Sigma(\omega)$ versus real frequencies, as obtained from Pad\'e approximants. It shows a finite imaginary part at $\omega=0$ indicating non-Fermi liquid behavior, where the real part shown in the inset has an inflection point. \textit{Bottom left panel} d) The finite imaginary part at zero frequency is confirmed by polynomial extrapolation with 9-th order polynomials.
\textit{Bottom mid panel} e) The double-log plot in the inset shows ${\rm Im}\Sigma$ for $N_W=16,\beta=30$ and interaction strengths $U=2.5-3.9$ and fortifies the assumption that ${\rm Im}\Sigma(i\nu_n) \propto (i\nu_n)^{\alpha}$. The data for $U=3.9$ deviates from this behavior and does not extrapolate to zero at zero frequency.
\textit{Bottom right panel} f) Power-law exponent $\alpha$ extracted from fits to ${\rm Im}\Sigma(i\nu_n)$ on the Matsubara axis show that the system leaves the Fermi-liquid regime as U increases with roughly a $\nu^{1/2}$ behavior along the critical line.}
\label{fig:qpw_alpha}
\end{figure*}

In order to check whether the system becomes insulating at finite temperatures we compute the quasiparticle renormalization factor according to
\begin{equation}\label{eq:zeee}
Z=\left[1-\frac{\partial {\rm Re} \Sigma_{\kay,\omega}}{\partial \omega}\right]^{-1}_{\kay=\kay_F,\omega=0},
\end{equation}
where $\omega$ labels real frequencies. We perform analytical continuation using Padé approximants. The result is plotted against the interaction strength in panels a) and b) of Fig.~\ref{fig:qpw_alpha}. Though initially decreasing as expected, $Z$ is seen to increase with $U$ above a certain threshold. The figure illustrates that this behavior occurs for different temperatures and number of bosonic frequencies alike. In the following we will show that the system undergoes a crossover to a non-Fermi liquid where the quasiparticle renormalization factor is ill defined and the unphysical increase of $Z$ with $U$ is  due to the inapplicability of Eq.~\eqref{eq:zeee}. For this purpose we analyze the self-energy at the Dirac point. In panel c) we show the result of the analytically continued self-energy for $U=3.8$ and $T=1/30$. The main panel shows the imaginary part, while the real part is shown in the inset. Note that $\mathrm{Im}\Sigma_{\mathbf{K},\omega=0} \neq 0$. The real part of the self-energy has an inflection point around $\omega=0$. The occurrence of a finite imaginary part of the self-energy at zero frequency is confirmed using polynomial extrapolation (the result is robust with respect to the order of the polynomials) as can be seen in panel d) of Fig.~\ref{fig:qpw_alpha}. In order to further quantify the crossover to the non-Fermi liquid we assume that the imaginary part of the self-energy on the Matsubara axis obeys a power law ${\rm Im}\Sigma\propto (\nu_{n})^{\alpha}$, as expected at the critical point. The doubly logarithmic plot shown in Fig.~\ref{fig:qpw_alpha} e) indicates that this is a reasonable assumption up to roughly $U=3.8$, where the self-energy visibly deviates from a power law. The exponent $\alpha$ is plotted against $U$ in Fig.~\ref{fig:qpw_alpha} f). As U increases, the value of $\alpha$ clearly approaches 1/2. 

Similarly to the observation in Refs.~\onlinecite{spinfreeze,spinfreeze_II}, the critical line separating the two phases is characterized by a square root behavior of the self-energy ${\rm Im}\Sigma\propto\nu^{1/2}$. We note that the self-energy on the real frequency axis in panel c) bears similarity to the results shown in Ref.~\onlinecite{spinfreeze}. However we are not able to establish the existence of frozen moments from the susceptibility data.

The finite imaginary part of the self-energy at the Fermi surface together with the exponent approaching $\alpha=1/2$ are clear signs of incipient non-Fermi liquid behavior. Note from Fig.~\ref{fig:qpw_alpha} a) and f) that the point where Z takes a minimum in fact agrees well with the point where $\alpha$ is expected to reach 1/2 and underlines that the result for $Z$ based on the Fermi-liquid formular~\eqref{eq:zeee} for $U>3.8$ is indeed unphysical.
Physically the finite imaginary part of the self-energy is due to scattering of electrons off spin fluctuations. Note that these spin fluctuations are associated with a magnetic response causing the ladder series to become unstable as the leading eigenvalue grows large, prohibiting us from accessing values of $U$ beyond the crossover point. We find that the ladder series becomes unstable when $1-\lambda \simeq 10^{-4}$. Therefore we cannot directly observe how the system turns insulating as the scattering increases. We find that the values of $U$ where the self-energy starts flattening out at low frequencies agrees well with the crossover to the renormalized classical regime (not shown). In Ref.~\onlinecite{wuandtremblay} the finite temperature Mott transition was preceded by a so called bad insulator, characterized by flattening of the self-energy at low Matsubara frequencies similar to our data.

\section{Conclusions}\label{sec:conc}
In summary, we have presented a multiband extension of the Dual Fermion approach and applied it to the half-filled Hubbard model on the honeycomb lattice. We have found clear signs of the quantum phase transition from a paramagnetic semimetal to an AFM insulator in our data, in particular in the size dependence and divergence of the leading eigenvalue of the Bethe-Salpeter equation as well as in the self-energy. Analysis of the leading eigenvalue has further enabled us to compute the crossover temperature $T_x$ to the renormalized classical regime and connect the strong-coupling limit to the quantum nonlinear $\sigma$ model. The numerical value where AFM ordering sets in at $U_c\approx 3.6-3.8$, is in good agreement with a variety of other numerical methods~\cite{Sorella,PhysRevX.3.031010,wuandtremblay,wuliebsch}. Analysis of the self-energy shows clear signs that for $U>U_c$ the system behaves like a non-Fermi liquid at finite temperature. 
In the accessible regime of the $U$-$T$ phase diagram we do not observe the semimetal-insulator transition. Extrapolation of the quasi-particle weight to zero temperature nevertheless suggests that the system becomes antiferromagnetic before turning insulating, which would exclude the existence of a spin-liquid phase.

\section{Acknowledgements}
The authors thank Eugene Kogan, Alexey N. Rubtsov, Mikhail I. Katsnelson and Guy Cohen for fruitful discussions. This work has been supported by Deutsche Forschungsgemeinschaft through the excellence cluster "The Hamburg Centre for Ultrafast Imaging - Structure, Dynamics and Control of Matter at the Atomic Scale" and from the European Graphene Flagship. Computational resources were provided by the HLRN-Cluster under Project No. hhp00030.

\appendix
\section*{Multiband Dual Fermion Approach}
We provide an outline of the derivation of the multiband formalism for local Coulomb interaction for the paper to be self-contained.

The partition function and action for the multiband Hubbard model are given by
\begin{align}
\mathcal{Z} &= \int \mathcal{D}[c\str,c] \exp{(-S[c\str,c])} \label{eq:A_hub_action_I}\\
S[c\str,c] = &-\sum_{\nu\kay AB\sigma} c\str_{\nu\kay A\sigma}\left[ (i\nu+\mu)\delta_{AB} -\hat{\epsilon}(\kay)_{AB} \right]c_{\nu\kay B\sigma} \notag \\ 
&+ U \sum_{\omega iA}n_{\omega i A \up} n_{-\omega i A \down}.\label{eq:A_hub_action_II}
\end{align}
The labeling conventions are the same as in the main text.
The main step is to express the lattice action in terms of the impurity action $S_\text{imp}$ by adding and subtracting an arbitrary, frequency dependent hybridization matrix $\hat{\Delta}_\nu$ to the action Eq.~\eqref{eq:hub_action}, 
\begin{equation}
S[c\str,c] = \sum_{i A} S_{\text{imp}}[c\str_{i A},c_{i A}]-\!\!\!\sum_{\nu\kay AB\sigma}\!\! c\str_{\nu\kay  A\sigma}\left[ \hat{\Delta}_{\nu}-\hat{\epsilon}_{\kay} \right]_{A B\sigma}c_{\nu\kay B}, \label{eq:A_hub_action_IV}
\end{equation}
where $S_{\text{imp}}[c\str_{i A},c_{i A}]$ is the action of a single impurity Anderson model on site $i$ and sublattice $A$:
\begin{align}
S_{\text{imp}}[c\str,c] &= -\sum_{\nu\sigma}c\str_{\nu\sigma}\left[ (i\nu+\mu)\mathbf{1} -\hat{\Delta}_{\nu}\right]_{AB}c_{\nu \sigma} + U \sum_{\omega}n_{\omega\up} n_{-\omega\down} \label{eq:A_imp_action}.
\end{align}
Dual fermions, represented by Grassmann fields $f$ and $f\str$, are introduced using a Hubbard-Stratonovich transformation (HST) to the second term in Eq.~\eqref{eq:A_hub_action_IV} according to
\begin{align}
\exp(c\str_{i}\hat{b}_{ij}\hat{a}^{-1}_{jk}\hat{b}_{kl}c_{l})&= \notag\\
\frac{1}{\det \hat{a}}\int \mathcal{D}[f\str,f] &\exp(-f\str_{i}a_{ij}f_{j}+f\str_{i}b_{ij}c_{j}+c\str_{i}b_{ij}f_{j}) \label{eq:A_HST_I} \\
\hat{a}&=\hat{g}^{-1}\left[\hat{\Delta}-\hat{\epsilon}_\kay \right]\hat{g}^{-1} \label{eq:A_HST_II}\\ 
\hat{b}&=-\hat{g}^{-1}\label{eq:A_HST_III},
\end{align}
where $g$ is the impurity Green's function corresponding to the impurity problem defined by Eq.\eqref{eq:A_imp_action} and $\hat{a}$ and $\hat{b}$ are in principle arbitrary matrices which are set to above expressions for convenience. Thus one arrives at
\begin{align}\label{eq:A_mixed_action}
S[c\str,c,f\str,f] &= \sum_{i} S_\text{site}[c\str_i,c_i] \notag \\
+&\sum_{\nu,\kay,A,B,\sigma} f\str_{\nu\kay A\sigma}\left[\hat{g}_{\nu}^{-1}\left(\hat{\Delta}_\nu-\hat{\epsilon}_{\kay}\right)^{-1}\hat{g}_{\nu}^{-1}\right]_{AB}f_{\nu\kay B\sigma} \\
S_\text{site}[c\str_i,c_i] &= S_{\text{imp}}[c\str_i,c_i] \notag \\
+&\sum_{\nu,i,A,B,\sigma} f\str_{\nu iA\sigma} \left[ \hat{g}_{\nu}\right]^{-1}_{AB}c_{\nu i B\sigma} + c\str_{\nu i A\sigma} \left[ \hat{g}_{\nu}\right]^{-1}_{AB}f_{\nu i B\sigma}.
\end{align}
In this form the lattice fermions, represented by $c$ and $c\str$, from different sites are decoupled and can be formally integrated out, as outlined below. Thereafter one is left with an action that is entirely formulated in terms of dual fermions 
\begin{align}
\tilde{S}[f^{*},f] &=-\sum_{\nu AB\sigma} f^{*}_{\nu\kay A\sigma}\left[\tilde{G}^{(0)}_{\nu\kay}\right]_{AB}^{-1}f_{\nu\kay B\sigma} + \sum_i \tilde{V}[f^{*}_i,f_i], \label{eq:A_dual_action} \\
\hat{\tilde{G}}^{(0)}_{\nu\kay} &=\left[\hat{g}^{-1}_\nu+(\hat{\Delta}_\nu-\hat{\epsilon}_\kay)\right]^{-1}-\hat{g}_\nu, \label{eq:A_bare_dual} \\
\tilde{G}_{\nu\kay AB} &= \frac{-1}{\tilde{\mathcal{Z}}}\int \mathcal{D}[f\str,f] f_{\nu\kay A} f\str_{\nu\kay B}\exp(-\tilde{S}[f\str,f]),
\end{align}
where $\hat{\tilde{G}}^{(0)}_{\bar{\nu}}$ is the bare propagator for dual fermions and the third line defines the dual Green's function. The dual partition function is given by $\tilde{\mathcal{Z}}=\mathcal{Z}/\det\left[{\hat{g}(\hat{\Delta}-\hat{\epsilon})\hat{g}}\right]$. 
The second term in Eq.~\eqref{eq:A_dual_action} gathers all terms that are of higher order in $f\str$ and $f$ than the bilinear term. It defines the interaction $V[f\str,f]$ between dual fermions and results from expanding $\exp(-S_\text{site})$ in powers of $f\str$ and $f$. Because the partition function contains $\exp(-S_\text{imp})$, integrating out the lattice fermions produces (connected) correlation functions of the impurity. The dual interaction is correspondingly given by
\begin{align}
V[f\str,f] &=-\frac{1}{4}\gamma^{A}_{\alpha\beta\gamma\delta}f\str_{A \alpha}f_{A \beta}f\str_{A \gamma}f_{A \delta} + \ldots \\
\gamma^{A}_{\alpha\beta\gamma\delta} &= g^{-1}_{A \alpha\alpha'} g^{-1}_{A \gamma\gamma'}\left[\chi ^\text{imp}_{A \alpha'\beta'\gamma'\delta'}-\chi^\text{imp,0}_{A \alpha'\beta'\gamma'\delta'} \right]g^{-1}_{A \beta'\beta} g^{-1}_{A \delta'\delta} \\
\chi^{\text{imp}}_{A \alpha\beta\gamma\delta}:&=\frac{1}{\mathcal{Z}_\text{imp}}\int c_{A\alpha} c\str_{A \beta} c_{A \gamma} c\str_{A \delta }\exp\left(-S_\text{imp}[c\str ,c]\right)\mathcal{D}[c\str ,c]\\
\chi ^\text{(0)imp}_{A \alpha\beta\gamma\delta}&=g_{A \alpha\beta}g_{A \gamma\delta }-g_{A \alpha\delta}g_{A \gamma\beta},
\end{align}
where all degrees of freedom are merged into a compound index $\alpha=\{\nu\sigma\}$. Repeated indices are summed over by convention. $\gamma$ is the reducible two-particle impurity vertex. $\chi_\text{imp}$ is the two-particle Green's function of the impurity and $\chi^{(0)}_\text{imp}$ is its disconnected part.

In order to compute observables of the original system consisting of lattice fermions $c\str$ and $c$, dual quantities have to be transformed after convergence of the perturbation series. In order to find the proper transformation consider the equality
\begin{align}
\mathcal{Z} &= \int \mathcal{D}[c\str,c] \exp{(-S[c\str,c])} \notag\\
&=\mathcal{Z}_f\int \mathcal{D}[c\str,c,f\str,f] \exp{(-S[c\str,c,f\str,f])},\label{eq:A_pf_I} \\
\mathcal{Z}_f&=\frac{\det\left[{\hat{g}(\hat{\Delta}-\hat{\epsilon})\hat{g}}\right]}{\prod_{iA}\mathcal{Z}^{iA}_{imp}},
\end{align}
which follows from the HST Eqs.~\eqref{eq:A_HST_I}-\eqref{eq:A_HST_II} and $\mathcal{Z}^{iA}_{imp}$ is the impurity located at site $i$ on sublattice $A$. Taking the functional derivative of Eq.~\eqref{eq:A_pf_I} with respect to the one-particle Hamiltonian yields an identity relating the Green's function of dual and lattice fermions
\begin{equation}\label{eq:A_dual_lattice_relationI}
\hat{G}=\left[\left(\hat{g}+\hat{g}\hat{\tilde{\Sigma}}\hat{g} \right)^{-1}+\hat{\Delta}-\hat{\epsilon} \right]^{-1}.
\end{equation}
Taking the according second derivatives yields a corresponding relation between two-particle Green's functions, which in turn can be used to derive a relation between the full two-particle vertices of dual and lattice fermions
\begin{align}
\Gamma_{\alpha\beta\gamma\delta} &=L'_{\alpha\alpha'}L'_{\gamma\gamma'}\tilde{\Gamma}_{\alpha'\beta'\gamma'\delta'}R'_{\beta'\beta}R'_{\gamma'\gamma} ,\notag\\
\hat{R}' &=\hat{\tilde{G}}\hat{g}^{-1}\left[\hat{\Delta}-\hat{\epsilon}\right]^{-1}\hat{G}^{-1},\notag\\
\hat{L}' &=\hat{G}^{-1}\left[\hat{\Delta}-\hat{\epsilon}\right]^{-1}\hat{g}^{-1}\hat{\tilde{G}}.
\label{eq:A_dual_vertex_relation}
\end{align}
This relation allows one to compute the generalized susceptibility tensor from the renormalized dual vertex and the dual Green's function on the fly by using it in Eq.~\eqref{eq:gen_susc_ten}.

\bibliographystyle{apsrev4-1}
\bibliography{main}
\end{document}